\documentclass[eclepsf,showpacs,preprintnumbers,amsmath,amssymb]{revtex4}
\usepackage{graphicx}

\def\title#1{\begin{center}{\Large\bf #1}\end{center}}
\def\author#1{\vskip 5mm \begin{center}{#1}\end{center}}
\def\address#1{\begin{center}{\it #1}\end{center}}
\newcommand{\simgt}{\lower.5ex\hbox{$\; \buildrel > \over \sim \;$}}
\newcommand{\simlt}{\lower.5ex\hbox{$\; \buildrel < \over \sim \;$}}

\begin{document}

\title{Dark energy reflections in the redshift-space quadrupole}

\author{Kazuhiro Yamamoto$^{1}$, Bruce A. Bassett$^{2,3}$ and Hiroaki Nishioka$^4$ 
}
\address{
$^1$Department of Physical Science, Hiroshima University,
Higashi-Hiroshima, 739-8526,~Japan\\ 
$^2$Department of Physics, Kyoto University,
Kyoto,739-8502,~ Japan\\
$^3$ ICG, University of Portsmouth, PO12EG, England\\
$^4$Institute of Astronomy and Astrophysics, 
Academia Sinica, Taipei 106, Taiwan, R.O.C.
}

\begin{abstract}
We show that  the redshift-space quadrupole will be a powerful tool  
for constraining dark energy even if the baryon oscillations are missing from the usual monopole 
power spectrum and bias is scale- and time-dependent. We calculate the accuracy with which a
next-generation galaxy survey based on  KAOS will measure the quadrupole power spectrum, 
which gives the leading anisotropies 
in the power spectrum in redshift space due to the linear velocity,  Finger of God  and Alcock-Paczynski effects. 
Combining the  monopole and quadrupole power spectra 
breaks the degeneracies between the multiple bias parameters and dark energy both in the linear and 
nonlinear regimes and, 
in the complete absence of baryon oscillations ($\Omega_b=0$), leads 
to a roughly 500\% improvement in constraints on dark energy compared 
with those from the monopole spectrum alone. 
As a result the worst case -- with no baryon oscillations -- has dark energy 
errors only mildly degraded relative to the ideal case, providing insurance 
on the robustness of  next-generation galaxy survey constraints on dark energy.
\end{abstract}
\maketitle
\def\M{{M}}
\def\w{{\psi}}
\def\calP{{\cal P}}

{\em Introduction}~~~The promise of next-generation galaxy surveys such as that planned with KAOS 
(the Kilo-Aperture Optical Spectrograph \cite{KAOS}) \footnote{In this paper, `KAOS' will refer to surveys performed using 
a KAOS-like wide field multi-object spectrograph. KAOS1 denotes a putative survey at $z<1.5$ while KAOS2 refers to a survey
at $2.5 < z < 3.5$. Since we never need to discuss the actual spectrograph itself this slight abuse of terminology, between the instrument and the survey conducted with that same instrument, should not cause confusion. } is to map the distribution 
of over one million galaxies in the redshift range $z= 0.5 - 3.5$. 
This redshift coverage will allow the baryon oscillations in the 
matter power spectrum to be followed as they were stretched by 
the cosmic expansion, thus providing us with a standard ruler 
with which to precisely measure the extragalactic distance 
scale and expansion rate \cite{eisen,Blake,Linder,SE,HH,Amendola}. 

However, this technique relies crucially on the assumption that 
the baryon oscillations will be detected. Although there 
are tentative indications for this at low-$z$ in the 2df data 
\cite{2dfosc,Yamamoto2004} the jury is still out on their existence. 
If bias turns out to be much more complicated than we think or 
$\Omega_b$ is unexpectedly low we may face an essentially 
featureless galaxy power spectrum that is too slippery to 
supply a standard ruler. In that case it is natural to ask 
whether surveys such as KAOS
will yield any constraints on dark energy at all.

The aim of this letter is to show that {\em even} in this worst case scenario, 
next-generation surveys will be able to deliver 
good constraints on dark energy through a very different route: redshift-space 
anisotropies and the Alcock-Paczynski (AP) effect  
\cite{Alcock,BPH,MS,MSz,matubara,Yamamoto2003}. 

In general the power spectrum {\em in redshift space} is not isotropic; an effect already seen in
the 2df survey \cite{Peacock}. There is a linear distortion due to the bulk motion of the sources within 
the linear theory of density perturbation \cite{Kaiser}, 
while the Finger of God effect causes radial elongations due to the 
motion of galaxies in the nonlinear regime \cite{PD}. 
In addition there is a geometric distortion due to the AP effect related to the 
distance-redshift relation of the universe. As a result the 
redshift-space power spectrum depends on the angle $\theta$ 
between the line-of-sight direction $\gamma$ and the wave 
number vector ${{\bf k}}$ (see e.g., \cite{SMJMY}). 

In general the redshift-space power spectrum can be expanded as \cite{TH,Hamilton}:
\begin{eqnarray}
  P({\bf k},z)=P(k,\mu,z)=\sum_{l=0,2,4\cdots} P_{l}(k,z){\cal L}_{l}(\mu),
\end{eqnarray}
where ${\cal L}_{l}(\mu)$ is the Legendre polynomial, 
$\mu=\cos\theta$ and $k=|{\bf k}|$. The odd moments vanish by symmetry.

The monopole  $P_{0}(k,z)$ represents the angular averaged power 
spectrum and is usually what we mean by `the power spectrum'. At low-z it has been investigated in great depth 
in the 2df and SDSS surveys.  $P_{2}(k,z)$ is the quadrupole spectrum and gives the leading anisotropic contribution. As can be 
seen in Fig.~1 it will be well-constrained even by just the $z<1.5$ sample, which we label KAOS1 
(see Table 1 for definitions). The higher order multipoles are not  well-constrained however.

Crucially, the multipole moments reflect different 
aspects of the redshift distortions in the power spectrum which can therefore aid in 
breaking degeneracies between the cosmological parameters, bias and dark energy.
The purpose of this {\em letter} is to consider the extent to which the anisotropic 
component of the power spectrum, $P_{\ell},~\ell \geq 2$, gives new information about dark 
energy via the nonlinear effects and the geometric (AP) distortion.

\def\dls{{D_{\rm LS}}}
\def\dos{{D_{\rm OS}}}
\def\bftheta{{\Theta}}
\def\calD{{\cal D}}
\def\bfk{{\bf k}}
\def\bfs{{\bf s}}

\begin{figure*}
\includegraphics[scale=0.8]{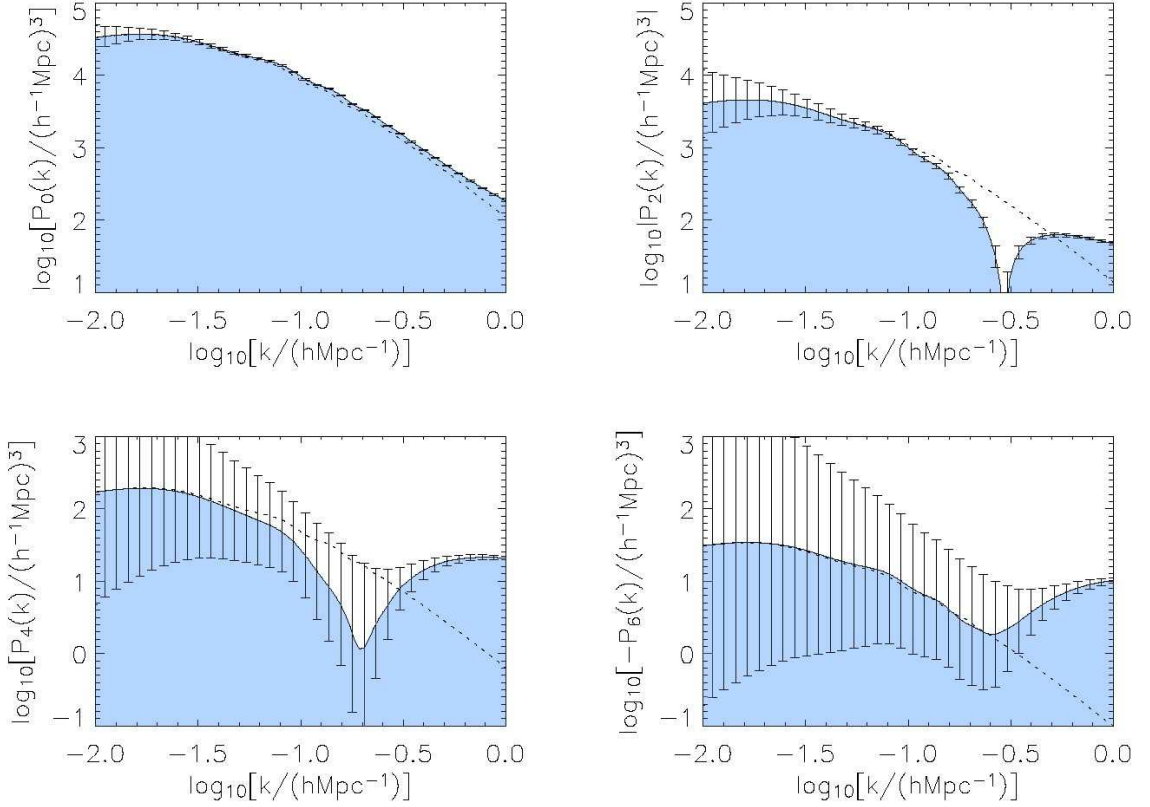}
\caption{\label{fig1} {\bf KAOS1 constraints ($z<1.5$) on the multipole moments of the power spectrum}, 
$\langle \calP_0(k)\rangle$, $\langle \calP_2(k)\rangle$, $\langle \calP_4(k)\rangle$ and $\langle \calP_6(k)\rangle$ 
for linear (dotted lines) and nonlinear (shaded region, solid lines) spectra respectively. While the nonlinear correction to $\calP_0$ 
(the usual `power spectrum') is small this is not true for the anisotropic spectra. The nonlinear
$\langle \calP_2(k)\rangle$ changes sign at large $k$. Here we have fixed $n=1$, $h=0.7$, 
$\Omega_b=0.045$,  $\Omega_m=0.28$ and $w=-1$. For the bias we adopted $b_0=1.35$, 
$p_0=1$ for the linear model, $p_1=1$, $b_1=0.1$, $\nu=1$ for the 
nonlinear spectrum (see eq. \ref{nlbias}).  The higher moments $\calP_{\ell},~~ \ell \geq 4$ are not 
well-constrained, even by KAOS and hence make a minimal contribution to constraints on dark energy (see Fig. ~2).
}
\end{figure*}

{\em Formalism}~~~Here we employ the Fisher matrix approach in order to estimate
the accuracy with which we can constrain the equation of state, $w \equiv p/\rho$,
of the dark energy with a measurement of the power 
spectrum. In general the Fisher matrix is defined by
$ { F}_{ij}=-\bigl\langle {\partial^2 \ln L
  /\partial\theta_i \partial\theta_j}
  \bigr\rangle$,
where $L$ is the likelihood of a data
set given the model parameters $\theta_i$. Assuming a Gaussian probability distribution function for 
the errors of a measurement of the multipole power spectrum 
${\calP}_{l}(k)$, the Fisher matrix for each multipole spectrum is
\begin{eqnarray}
  F_{ij}^{(l)}\simeq
  {1\over 4\pi} \int_{k_{\rm min}}^{k_{\rm max}} \kappa_{l} (k)
  {\partial \langle{\calP}_{l}(k)\rangle\over \partial \theta_i}
  {\partial \langle{\calP}_{l}(k)\rangle\over \partial \theta_j}
  k^3 d\ln k,
\end{eqnarray}
where $\kappa_{l}(k)$ is the effective volume of the survey available for measuring 
$\cal P_{\ell}$ at wavenumber $k$:
\begin{eqnarray}
 \kappa_{l}(k)^{-1}=
    {1\over 2}
  \int_{-1}^1 d\mu  
   { \int d\bfs \bar n(\bfs)^4 \w(\bfs,k,\mu)^{4}
     \bigl[P(k,\mu,z)+1/\bar n(\bfs)\bigr]^2
  [{\cal L}_{l}(\mu)]^2
  \over
  \Bigl[\int d\bfs' \bar n(\bfs')^2 \w(\bfs',k,\mu)^{2}\Bigr]^2},
\label{optimal}
\end{eqnarray}
and  
\begin{eqnarray}
  \langle {\calP}_{l}(k)\rangle=
  {1\over 2}\int_{-1}^1d\mu
  {\int d\bfs \bar n(\bfs)^2\w(\bfs,k,\mu)^2 P(k,\mu,z)
  {\cal L}_{l}(\mu)
\over
  \int d\bfs' \bar n^2(\bfs') \w(\bfs',k,\mu)^2},
\end{eqnarray}
where $\w(\bfs,k,\mu)$ is a weight factor that we can choose freely,
$\bar n(\bfs)$ is the mean number density, and $\bfs$ 
denotes the three dimensional coordinate in redshift space.  
This formula can be derived in a similar way to obtain the optimal 
weighting scheme (see e.g., \cite{FKP,Yamamoto2003}).
Minimizing the variance on the power spectrum yields 
 $ \w(\bfs,k,\mu)=[1+\bar n(\bfs) P(k,\mu,z)]^{-1}$,
the same as used in \cite{SE}. 

\newcommand{\cpara}{c_{\scriptscriptstyle \|}}
\newcommand{\cperp}{c_{\scriptscriptstyle \bot}}
\newcommand{\qpara}{q_{\scriptscriptstyle \|}}
\newcommand{\qperp}{q_{\scriptscriptstyle \bot}}

Next we explain our theoretical modelling of the power 
spectrum. In a redshift survey, the redshift $z$ is the indicator of the
distance. Therefore we need to assume a distance-redshift relation
$s=|\bfs|=s[z]$ to plot a map of objects. The power spectrum depends on 
this choice of the radial coordinate of the map $s=s[z]$ due to
the geometric distortion (AP) effect.  For our fiducial background we adopt  
a flat universe with $\Omega_m=0.3$. Here $H_0=100~h{\rm km/s/Mpc}$ is the Hubble parameter.
We consider a cosmological model with the dark energy 
component with constant equation of state, $w \equiv p/\rho$, since estimates for the nonlinear power 
spectrum in more general cases do not yet exist. While such an approach has severe limitations when 
extracting accurate conclusions from real data \cite{param}, it suffices for our purposes since we are mainly 
interested in understanding the qualitative improvements in the constraints on $w$ from inclusion of the 
quadrupole, especially as the baryon oscillations disappear from the monopole. 

For constant $w$ we have 
\begin{eqnarray}
  &&r(z,\Omega_m,w)={1\over H_0}\int_0^z{dz'\over
  \sqrt{\Omega_m(1+z')^3+(1-\Omega_m) (1+z')^{-3(1+w)}}}.
\label{defr}
\end{eqnarray}

Our fiducial model thus has $s(z) \equiv r(z,0.3,-1)$.
The geometric distortion in the power spectrum depends on $r(z,\Omega_m,w)$ 
and the power spectrum at redshift $z$ is described by 
scaling the wave numbers from real space to redshift space via 
$\qpara\rightarrow{k\mu/\cpara}$ and $\qperp\rightarrow
{k\sqrt{1-\mu^2}/\cperp}$ with 
$\cpara(z)={dr(z)/ds(z)}$ and $\cperp(z)={ r(z)/s(z)}$.
\begin{figure*}
\includegraphics[scale=0.8]{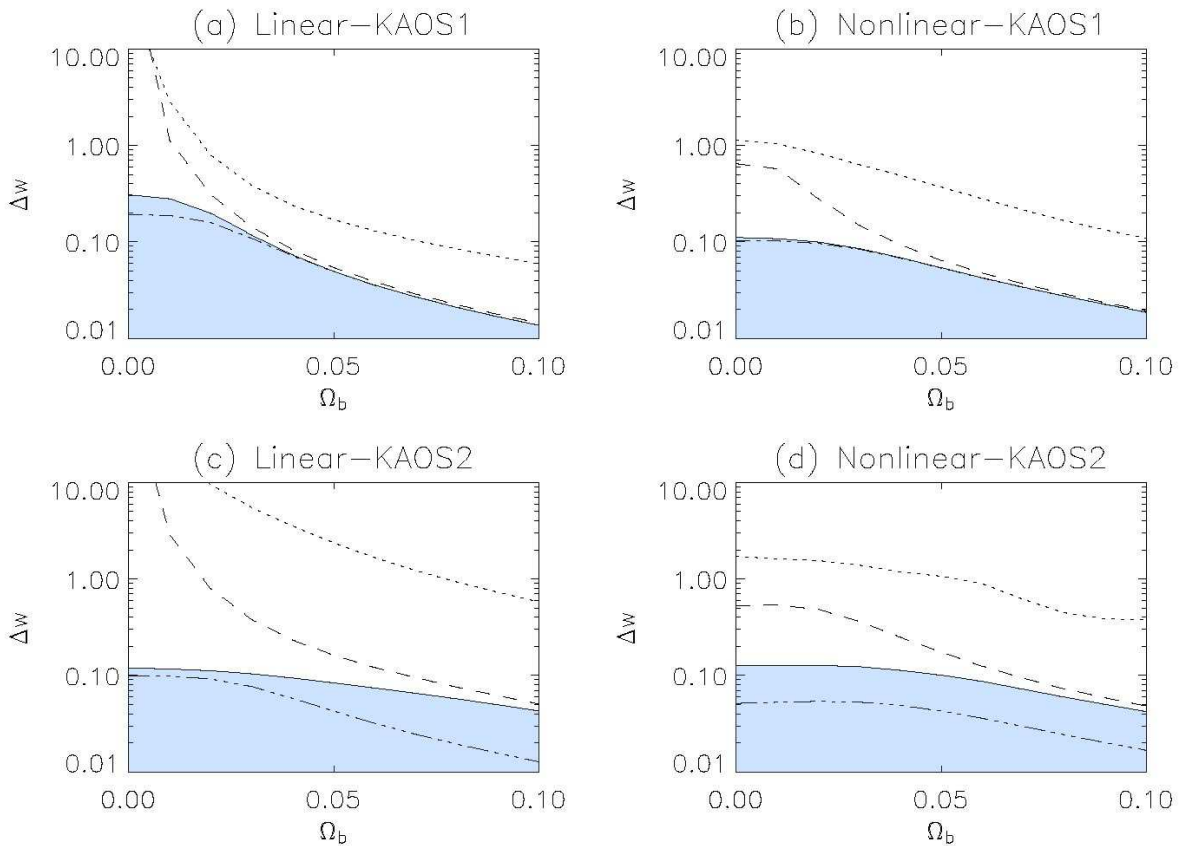}
\caption{\label{fig2} {\bf Error estimates for $w$ as the baryon oscillations disappear ($\Omega_b \rightarrow 0$)}.
The left panels (a) and (c) are the results using the linear spectrum and the right
panels (b) and (d) are the nonlinear spectrum. The dashed curve is the result
utilizing {\em only} $\calP_0(k)$, the dotted curve is the result 
with {\em only} $\calP_2(k)$, the solid curve (delimiting the shaded region) is the result obtained using
{\em both} $\calP_0(k)$ and $\calP_2(k)$. The target parameters here are same as those in Fig.~1.
The low-redshift sample, KAOS1, is assumed in (a) and (b), the high-redshift sample, KAOS2, 
is assumed in (c) and (d). The dotted-dashed curve in (a) and (b) shows the constraint
combining all $\calP_0(k)$ to $\calP_6(k)$. 
The double dotted-dashed curve in (c) and (d) shows the constraint
obtained from the full KAOS sample (KAOS1 + KAOS2). The key point is 
how flat the resulting curve is for $\Omega_b \leq 0.05$ despite the absence of baryon oscillations 
for $\Omega_b \rightarrow 0$. 
}
\end{figure*}

We write the galaxy power spectrum in nonlinear theory as
\begin{eqnarray}
  P_{\rm gal}(\qpara,\qperp,z)=
  \biggl(1+{f(z)\over b(z,q)}{\qpara^2\over q^2}\biggr)^2
  b(z,k)^2 P^{\rm NL}_{\rm mass}(q,z) D[\qpara],
\label{PQSO}
\end{eqnarray}
with
$f(z)=d\ln D_1(z)/d\ln a(z)$, where $q^2=\qpara^2+\qperp^2$, $b(z,q)$ is a scale-dependent
bias factor, $P^{\rm NL}_{\rm mass}(q,z)$ is the nonlinear 
mass power spectrum normalized by $\sigma_8=0.9$, 
$D_1(z)$ is the linear growth rate, and 
$a(z)$ is the scale factor.
The term in proportion to $f(z)$ describes the linear 
distortion \cite{Kaiser}.
$D[\qpara]$ represents the
damping factor due to the Finger of God effect. Assuming 
an exponential distribution function for the pair-wise 
peculiar velocity \cite{MJB,MJS,SMY} gives 
$  D[\qpara]=1/( 1+(\qpara\sigma_P)^2/2)$,
where $\sigma_P$ is the 1-dimensional pair-wise peculiar 
velocity dispersion estimated in  \cite{MJB}.

For  $P^{\rm NL}_{\rm mass}(q,z)$ we adopt the 
fitting formula for the quintessence cosmological 
model \cite{Ma}.

We then use the fitting formula for $f(z)$ developed in \cite{WS}. 
For the nonlinear modelling, we assume a four-parameter, scale-dependent, bias model
\begin{eqnarray}
 b(z,k)=\left(1+{b_0-1\over D_1(z)}\right)\left[
 1+{b_1}\left({D_1(z)^{p_1}q\over0.1~h{\rm Mpc^{-1}}}\right)^{\nu}
\right],
\label{nlbias}
\end{eqnarray}
where $b_0$, $b_1$, $p_1$ and $\nu$ are the nonlinear bias constants. In the linear case the bias is scale-{\em independent} and given by 
$b(z)=1+(b_0-1)D_1(z)^{-p_0}$ where $p_0$ is a constant.

{\em Results}~~~Fig.~1 shows the  power
 spectra $\langle \calP_0(k)\rangle$ (usual monopole), $\langle \calP_2(k)\rangle$ (quadrupole),
$\langle \calP_4(k)\rangle$ and  $\langle \calP_6(k)\rangle$ 
for the linear and nonlinear  models described above, assuming the  KAOS1 sample described in Table 1. 
$\langle \calP_0(k)\rangle$ is positive while the nonlinear effects 
cause $\langle \calP_2(k)\rangle$ to change sign at large $k$. 
For $\langle\calP_0(k)\rangle$, the linear power spectrum agrees well with 
the nonlinear power spectrum because  the two nonlinear contributions  to it
cancel out:  the Finger of God effect decreases the amplitude while
$P^{\rm NL}_{\rm mass}(k)$ increases the amplitude due to the nonlinearity at large $k$. 
By comparison, it is very clear that the linear theory is not good for the 
higher multipole moments on small scales, $k\simgt 0.1 h{\rm Mpc}^{-1}$.

\begin{figure*}
\includegraphics[scale=0.7]{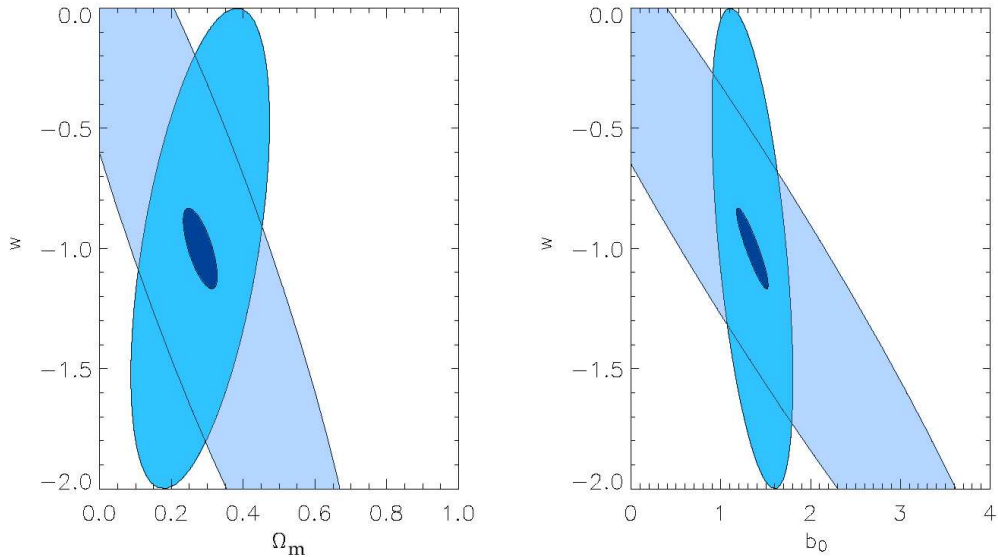}
\caption{\label{fig3}  {\bf The effect of adding the quadrupole}. 
 the $w-\Omega_m$ (left) and $w-b_0$ (right) likelihoods 
for $\calP_0$ alone (large, medium-blue ellipses),  
$\calP_2$ alone (largest, lightly-shaded ellipses) and $\calP_0 + \calP_2$ (small, 
darkest ellipses).  Here the same parameters are used as in Fig. 1 except that $\Omega_b=0$ so 
there are no baryon oscillations, which explains the poor constraints from $\calP_0$ alone. 
Adding the quadrupole significantly reduces the uncertainty in all the bias parameters 
resulting in final error ellipses that are competitive with those from the ideal 
baryon oscillation case, see Fig. 2. 
}
\end{figure*}
While the quadrupole will make a key contribution to constraining dark energy with 
KAOS,  the errors on $\langle \calP_4(k)\rangle$ and 
$\langle \calP_6(k)\rangle$ are large and hence their contribution to constraints 
on dark energy are marginal, as can be seen from the 
dot-dashed curves in panel (a) and (b) in Fig.~2.

\begin{table}
\begin{center}
\begin{tabular}{lcc}
\hline
~  & ~~~~~KAOS1~~~~~~ & ~~~~~KAOS2~~~~ \\
\hline
redshift range & $0.5 <z< 1.5$ & $2.5 <z< 3.5$ \\
survey area (deg${}^2$) & $103$ & $150$ \\
$\bar n~  ({\rm h^{3}Mpc^{-3}})$  & $10^{-4} $ & $10^{-4} $\\
$k_{\rm max} ~(h{\rm Mpc}^{-1})$  & $0.4 $ & $1 $ \\
$b_0$  &$1.35$ & $1.75$ \\
\hline
\end{tabular}
\end{center}
\caption{The parameters of the samples used in our analysis. 
$\bar n$ is the average galaxy number density, $k_{\rm max}$ is the 
maximum wavenumber used in evaluating the Fisher matrix.}
\end{table}

The precision with which $w$ can be recovered is shown in Fig 2. We 
consider separately the low-redshift, $z < 1.5$, (KAOS1) and high-redshift (KAOS2) samples, with 
parameters summarized in  Table 1. 

To produce these estimates we quote $\Delta w \equiv ({\bf F}^{-1/2})_{ww}$, marginalizing over $\Omega_m$ and 
all the bias parameters, {\em viz}  $b_0$, $p_0$ (linear case)  and 
$b_0$, $p_1$, $b_1$ and $\nu$ (nonlinear case). 
Since the bias may be constrained by other methods 
(e.g. lensing or the higher-order correlation function) our results 
are conservative. 

Fig.~2 shows $\Delta w$ as a function of $\Omega_b$.
The left panels are the results for the linear perturbation theory, 
while the right panels are the nonlinear model. 
The upper panels assume the KAOS1 sample, while the lower
panels assume the KAOS2 sample. 
In general, $\Delta w$ becomes larger as the baryon fraction 
becomes smaller since the baryon oscillations become less and 
less distinct. 
As $\Omega_b$ becomes smaller, the contribution from $\calP_2$ becomes 
increasingly important. It is clear from the dashed curve that 
the constraint on $w$ from $\calP_0$ is very weak around $\Omega_b=0$ because
the baryon oscillations disappear, taking with it the standard ruler.
This is the same for the dotted curve which shows the constraints from $\calP_2$. 

One of the main results of this paper is the solid curve 
which shows the constraint from the combination
of $\calP_0$ and $\calP_2$. It is good even in the case $\Omega_b=0$ when 
the baryon oscillation are missing, implying that 
the geometric distortion (AP test) plays the central role in 
constraining $w$. This does not depends on the bias parameters and 
inclusion of a constant parameter for stochastic bias does not alter our 
results \cite{TP}.

It is interesting to address why the constraint on $w$ 
from $\calP_0$ and $\calP_2$ combined is so much better than from either one 
separately.  For each pair of marginalized parameters the error ellipses for $\calP_0$ and $\calP_2$ are
rotated with respect to each other, as in Fig~3, thus breaking degeneracies in the bias-dark 
energy parameter space. On marginalization these gains are passed through to $w$, resulting in significantly smaller 
error-ellipses, a feature observed in both the linear and nonlinear cases.  
The power in combining $\calP_0$ and $\calP_2$ thus
extends the well-known fact that $\calP_2$ gives useful information 
about bias (e.g., \cite{TH,Hamilton}).

{\em Conclusions}~~~ We have investigated the accuracy with which we can expect next-generation 
galaxy surveys such as KAOS \cite{KAOS} to measure the  multipole moments of the anisotropic power spectrum in 
redshift space and the resulting improvements in dark energy constraints. Anisotropies in the redshift-space power spectrum 
arise from the contribution of velocities to an object's redshift as well as from the geometric distortion due to the 
Alock-Paczynski effect. 

We found a number of key results: (1)  only the quadrupole among the anisotropic power spectra 
will be well-measured by KAOS but this is useful for breaking 
degeneracies between bias and dark energy.  (2) Nonlinear effects have a substantial influence on the 
quadrupole and higher multipoles at scales 
$k\simgt 0.1 h{\rm Mpc}^{-1}$. The inclusion of the nonlinear power spectrum 
enhances the precision with which the dark energy can be constrained because 
the nonlinear effects increase the power at small scale which is also where 
constraints are good. 
The nonlinear regime provides us with new information about the dark energy, 
as has been discussed in different contexts (e.g., \cite{ND}).
(3) Applying these results to dark energy and the KAOS survey we have found that significant constraints 
arise by combining the monopole and quadrupole spectra even if there are {\em no baryon oscillations} in the standard 
monopole spectrum  and even if we allow for multi-parameter scale-dependent or stochastic bias. 

This is a key piece of insurance for large galaxy surveys given 
current uncertainty about the existence of baryon oscillations 
and ensures that large, next-generation, 
galaxy surveys will make a significant contribution to the hunt 
for dark energy irrespective of the existence of baryon oscillations.

{\em Acknowledgements}~~~We thank Daniel Eisenstein and Bob Nichol for comments on the draft. 
This work is supported by Grant-in-Aid for  Scientific Research of Japanese Ministry of Education, 
Culture, Sports, Science, and Technology 15740155 and by a Royal Society/JSPS 
Fellowship.




\end{document}